\documentclass[12pt,twoside]{article}
\usepackage{amsmath,amssymb}

\newcommand{\cL}{{\cal L}}

\newcommand{\cM}{{\cal M}}
\newcommand{\cD}{{\cal D}}

\newcommand{\cB}{{\cal B}}

\newcommand{\cT}{{\cal T}}
\def\R{{\mathbb R}}

\newcommand{\ch}{{\mathfrak h}}

\newtheorem{thm}{Theorem}[section]

\newtheorem{prop}{Proposition}[section]
\newtheorem{lemm}{Lemma}[section]

\newtheorem{defn}{Definition}[section]
\newtheorem{rem}{Remark}[section]
\newtheorem{ass}{Assumption}[section]
\newtheorem{exam}{Example}[section]

\makeatletter

                    {$\square$\bigskip}

\@addtoreset{equation}{section}

\begin{document}

\title{}
\author{}
\date{}
\maketitle
\vspace{-5cm}

\hspace{0.1cm} \vspace{0.5cm} \baselineskip=18pt

\begin{center}

\noindent{\Large {\bf Quantum dynamical semigroups generated by
noncommutative unbounded elliptic operators }}
\end{center}

\begin{center}
\parbox{5in}
\noindent Changsoo Bahn

\noindent{\small  Natural Science Research Institute, Yonsei University, Seoul 120-749, Korea \\[-0.1cm]
\small e-mail: bahn@yonsei.ac.kr}\vskip 0.5 true cm

\noindent Chul Ki Ko

\noindent {\small Natural Science Research Institute, Yonsei University, Seoul 120-749, Korea\\[-0.1cm]
e-mail: kochulki@hotmail.com } \vskip 0.5 true cm

 \noindent Yong Moon Park

\noindent {\small Department of Mathematics, Yonsei University, Seoul 120-749, Korea\\[-0.1cm]
e-mail: ympark@yonsei.ac.kr} \vskip 0.2 true cm
\end{center}
\begin{abstract} We study quantum dynamical semigroups generated by noncommutative unbounded elliptic operators
which can be written as Lindblad type unbounded generators. Under
appropriate conditions, we first construct  the minimal quantum
dynamical semigroups for the generators and then use Chebotarev
and Fagnola's sufficient conditions for conservativity to show
that the semigroups are conservative.

\vspace*{0.3cm} \noindent
\end{abstract}

 \maketitle
 \section{Introduction}
\quad The purpose of this work is to study quantum dynamical
semigroups(q.d.s.) generated by noncommutative unbounded elliptic
operators. The generators can be expressed as Lindblad type
(unbounded) generators. Under appropriated conditions on
coefficients, we first construct the minimal quantum dynamical
semigroups for the generators and then use Chebotarev and
Fagnola's sufficient conditions for conservativity to show that
the semigroups are conservative.  For the details, see Section 3.

Let us first describe briefly the background of this study. In
\cite{BP}, using a quantum version of Feynman-Kac formula, the
authors constructed the Markovian semigroup generated by the
following noncommutative elliptic operator $\mathcal{L}$ on a von
Neumann algebra $\cM$ acting on a separable Hilbert space $\ch$:
\begin{align}\label{1.1}
 D(\mathcal{L})&=D(\delta^2) {,}\\
\mathcal{L}(X)&=\frac12
\delta^2(X)+a\delta(X)+\delta(X)a-\frac12[a,[a,X]],\,\,\,X\in
D(\mathcal{L}),\nonumber
\end{align}
where $a$ is a self-adjoint element of $\cM$, $\delta$ is the
generator of a weak*- continuous group of *-automorphisms
$(\alpha_t)_{t\in \R}$ of $\cM$ and $[A,B]=AB-BA$.

Let $\cM=B(\ch)$ and $b$ be a self-adjoint operator on $\ch$. Let
$\alpha_t(X)=e^{itb}Xe^{-itb},\,X\in\cM$, be the corresponding one
parameter group of automorphisms of $\cM$. Then
\begin{equation}\label{1.3}
\delta(X)=i[b,X],\,\,X\in D(\delta).
\end{equation}
 Put
 \begin{equation}\label{1-3}
L:= a-ib , \quad H:=\frac 12 (ab+ba).
\end{equation}
The generator $\cL$ in (\ref{1.1}) can be represented by the
following Lindblad type generator:
\begin{equation}\label{1.4}
\mathcal{L}(X)=i[H,X]-\frac12 L^*LX+L^*XL-\frac12 XL^*L,\,\,\,X\in
D(\mathcal{L}),
\end{equation}
where $[A,B]=AB-BA.$

 In this paper, we consider the following situation: Let $\ch=L^2(\R^d)$ and $W_l(x_1,\cdots,x_d)$, $l=1,2,\cdots,d$,
denoted by $W_l(x)$, be real valued twice differentiable functions
on $\R^d$. For each $l=1,2,\cdots,d$, let $\partial_l$ be the
differential operator $\frac{\partial}{\partial x_l}$ with respect
to the $l$-th coordinate. For each $l=1,2,\cdots, d,$ we choose
\begin{equation}\label{1.4-1}
 a_l =-W_l \,\, \quad  \text{and}\,\, \quad b_l =-i\partial_l.
\end{equation}
Then by (\ref{1-3})
$$
L_l :=-(W_l +\partial_l)  \,\,\quad  \text{and} \,\,\quad  H_l :=
\frac {i}{2} (W_l\partial_l + \partial_l W_l).
$$
We are interested in the following (formal) generater $\cL$ :
\begin{eqnarray}
\mathcal{{L}}(X) &=& \sum_{l=1} ^d \left( i[H_l,X]-\frac12
L_l^*L_lX+L_l^*X L_l-\frac12 XL_l^*L_l \right)\label{*1.4}\\
   &=& \sum_{l=1}^d \left( \frac12 [\partial_l, [\partial_l , X]] -W_l [\partial_l , X]
   -[\partial_l, X] W_l -\frac 12 [W_l, [W_l, X]] \right).\nonumber
\end{eqnarray}
It is worth to mention that if $X$ is a smooth function with a
compact support on $\R^d$ (a multiplication operator on $L^2
(\R^d)$), then $[W_l , X]=0$, $l=1,2,\cdots, d$ and the generator
given in (\ref{*1.4}) can be rewritten as
\begin{equation} \label{*1.5}
 \cL(X)
= \frac12 \Delta X - 2 W\cdot \nabla X,
\end{equation}
where $W=(W_1, \cdots , W_d)$, $\nabla X = (\partial_1 X, \cdots ,
\partial_d X)$ and $\Delta X = \sum_{l=1} ^d
\partial_{ll} X$. Thus the operator $\mathcal{{L}}$ given in
(\ref{*1.4}) is a noncommutaive generalization of the elliptic
operator given in (\ref{*1.5}).

The aim of this paper is to construct the conservative minimal
q.d.s. with generator $\cL$ given in (\ref{*1.4}) for an unbounded
multiplication operator $W_l$, $l=1,2,\cdots, d$. Because of the
unboundedness, the method of the quantum Feynman-Kac formula in
\cite{BP,LS} can not be applied. In \cite{BK}, the authors
employed the theory of the minimal quantum dynamical semigroup to
 construct the Markovian semigroup with generator $\cL$ in
(\ref{1.4}) under  the  condition $[a, b]$ is bounded. This
condition means that $[W_l,i\partial_l]$ is bounded for any
$l=1,2,\cdots, d$ in our case. In this paper, we will improve the
condition. Suppose that there exist positive constants $k_1$ and
$k_2$ such that the bounds
\begin{equation}\label{0-0}
|\frac{\partial W^2_l}{\partial x_k}|\le
k_1(W_1^2+...+W_d^2)+k_2,\quad l, k=1,2,...,d,
 \end{equation}
 hold(see Assumption \ref{ass3-0}).
Under additional conditions ( see Assumption \ref{ass3-0} and
Assumption \ref{ass3-1}), we construct the minimal q.d.s. with
generator $\cL$ given by (\ref{*1.4}) and show its conservativity
by using the result of Fagnola and Chebotarev\cite{CF1,CF2}.

The paper is organized as follows: In section 2, we review the
theory of the minimal q.d.s. and give Chebotarev and Fagnola's
sufficient conditions for conservativity \cite{CF2}. In section 3,
we give the assumptions and example for $W$ and state main
results. First, we introduce a proposition related to the
perturbation of generator of a strongly continuous contraction
semigroup, and then construct the minimal q.d.s. with (formal)
generater $\cL$. Under additional condition, we show that the
q.d.s. is conservative. Section 4 is devoted to proofs of main
results.

\section{Review on the minimal quantum dynamical semigroups}
\quad Let $\ch$ be a separable Hilbert space with the scalar
product $\langle\cdot,\cdot\rangle$ and norm $\|\cdot\|$. Let
$\mathcal{B}(\ch)$ denote the Banach space of bounded linear
operators on $\ch$. The uniform norm in $\mathcal{B}(\ch)$ is
denoted by $\|\cdot\|_\infty$ and the identity in $\ch$ is denoted
by $I$. We denote by $D(G)$ the domain of operator $G$ in $\ch$.
\begin{defn}
A quantum dynamical semigroup(q.d.s.) on $\mathcal{B}(\ch)$ is a
family $\mathcal{T}=(\mathcal{T}_t)_{t\ge0}$ of bounded operators
in $\mathcal{B}(\ch)$ with the following properties:
\begin{enumerate}
\item[(i)] $\mathcal{T}_0(X)=X,$ for all $X\in\mathcal{B}(\ch),$
\item[(ii)]$\mathcal{T}_{t+s}(X)=\mathcal{T}_{t}(\mathcal{T}_{s}(X)),$
for all $s,t\ge0$ and all $X\in\mathcal{B}(\ch),$
\item[(iii)]$\mathcal{T}_t(I)\le I,$ for all $t\ge0,$
\item[(iv)](completely positivity) for all $t\ge0$, all integers
$n$ and all finite sequences $(X_j)_{j=1}^n,\,(Y_l)_{l=1}^n$ of
elements of $\mathcal{B}(\ch)$, we have $$\sum_{j,\,l=1}^n
Y_l^*\mathcal{T}_t(X_l^*X_j)Y_j\ge0,$$
\item[(v)] (normality or $\sigma$-weak continuity) for every sequence $(X_n)_{n\ge1}$ of elements of $\mathcal{B}(\ch)$ converging
weakly to an element $X$ of $\mathcal{B}(\ch)$ the sequence
$(\mathcal{T}_t(X_n))_{n\ge1}$ converges weakly to
$\mathcal{T}_t(X)$ for all $t\ge0$,
 \item[(vi)] (ultraweak or weak\,$^*$ continuity) for
all trace class operator $\rho$ on $\ch$ and all
$X\in\mathcal{B}(\ch)$ we have
$$\lim_{t\rightarrow0^+}
Tr(\rho\mathcal{T}_t(X))=Tr(\rho X).$$
\end{enumerate}
\end{defn}
We recall that  as a consequence of properties (iii), (iv),  for
each  $t\ge0$ and $X \in \cB(\ch)$, $ \cT_t$ is
 a contraction, i.e.,
 \begin{equation}\label{2.0}
 \| \mathcal{T}_{t}(X)\|_{\infty}\le\|X\|_{\infty},
 \end{equation}
 and as a consequence of properties (iv), (vi), for all $X\in\mathcal{B}(\ch)$, the map
  $t\mapsto \mathcal{T}_{t}(X)$ is strongly continuous.

 \begin{defn}
A q.d.s. $\mathcal{T}=(\mathcal{T}_t)_{t\ge0}$ is called to be
conservative or Markovian if $\mathcal{T}_{t}(I)=I$ for all
$t\ge0$.
\end{defn}

The natural generator of q.d.s. would be the Lindblad type
generator\cite{Li,Par}
$$\mathcal{L}(X)=i[H,X]-\frac12 XM+\sum_{l=1}^{\infty} L_l^*XL_l-\frac12 MX,\quad X\in B(\ch)$$
where $M=\sum_{l=1}^{\infty} L_l^*L_l$, $L_l$ is densely defined
and $H$ a symmetric operator on $\ch$. The generator can be
formally written by
$$ \cL(X) =XG + G^* X + \sum_{l=1}^\infty
L_l^* X L_l,$$ where $G=-iH -\frac12 M.$ A very large class of
q.d.s. was constructed by Davies\cite{Da} satisfying the following
assumption. It is basically corresponding to the condition $\cL(I)
=0$.

\begin{ass}\label{ass2.1} The operator $G$ is the
infinitesimal generator of a strongly continuous contraction
semigroup $P=(P(t))_{t\ge 0} $ in $\ch$.  The domain of the
operators $(L_l )_{l=1} ^\infty $ contains the domain $D(G)$
 of $G$. For all $v, u \in D(G)$, we have
\begin{equation}\label{2.1}
\langle v,Gu \rangle+\langle Gv,u \rangle+\sum_{l=1}^\infty
\langle L_l v,L_l u\rangle=0.
\end{equation}
\end{ass}

 As a result of Proposition 2.5 of \cite{CF1} we can assume only that
the domain of the operators $L_l$ contains a subspace $D$ which is
a core for $G$ and (\ref{2.1}) holds for all $v,u\in D$.

For all $X\in \mathcal{B}(\ch)$, consider the sesquilinear form
$\mathcal{L}(X)$ on $\ch$ with domain $D(G)\times D(G)$ given by
\begin{equation}\label{2.2}
\langle v,\mathcal{L}(X)u\rangle=\langle v,XGu \rangle+\langle
Gv,Xu \rangle+\sum_{l=1}^\infty\langle L_l v,XL_l u\rangle.
\end{equation}
 Under the Assumption \ref{ass2.1} one can construct a q.d.s.
$\cT=(\cT_t)_{t\ge0}$ satisfying the equation
 \begin{equation}\label{2.3}
\langle v,\cT_t(X)u \rangle=\langle v,Xu \rangle+\int_0^t\langle
v,\mathcal{L}(\cT_s(X))u\rangle ds
\end{equation}
 for all $v ,u\in D(G)$ and all $X\in B(\ch)$. Indeed, for a strongly continuous family $(\cT_t(X))_{t\ge0}$ of
elements of $\mathcal{B}(\ch)$ satisfying (\ref{2.0}), the
followings are equivalent:
\begin{enumerate}
\item[(i)] equation (\ref{2.3}) holds for all $v ,u\in D(G)$,
\item[(ii)] for all $v ,u\in D(G)$ we have
\begin{align}\label{2.4}
\langle v,\cT_t(X)u \rangle&=\langle P(t) v, XP(t) u
\rangle\\&\quad+\sum_{l=1}^\infty\int_0^t\langle L_lP(t-s)
v,\cT_s(X)L_l P(t-s) u\rangle ds.\nonumber
\end{align}
\end{enumerate}
We refer to the proof of Proposition 2.3 in \cite{CF2}.
 A solution of the equation (\ref{2.4}) is
obtained by the iterations
\begin{align}\label{2.5}
\langle u,\cT_t^{(0)}(X)u\rangle&=\langle
P(t)u,XP(t)u\rangle,\\
\langle u,\cT_t^{(n+1)}(X)u\rangle&=\langle P(t)u,XP(t)u\rangle\nonumber\\
&\quad+\sum_{l=1}^\infty\int_0^t \langle
L_lP(t-s)u,\cT_s^{(n)}(X)L_lP(t-s)u\rangle ds \nonumber
\end{align}
for all $u\in D(G)$. In fact, for all positive elements
$X\in\mathcal{B}(\ch)$ and all  $t\ge0$, the sequence of operators
$(\cT_t^{(n)}(X))_{n\ge0}$ is non-decreasing. Therefore it is
strongly convergent and  its limits for $X\in\mathcal{B}(\ch)$ and
$t\ge0$ define the {\it minimal solution} $(\cT_t)_{t\ge0}$ of
(\ref{2.4}) in the sense that, given another solution
$(\cT_t')_{t\ge0}$ of (\ref{2.3}), one can easily check that
$$\cT_t(X)\le \cT_t'(X)\le \|X\|_\infty  I$$ for any
positive element $X$ and all $t\ge0$. For details, we refer to
\cite{Ch1,Fa}. From now on, the minimal solution $(\cT_t)_{t\ge0}$
is called the minimal q.d.s..

Chebotarev and Fagnola gave a criteria to verify the
conservativity of minimal q.d.s. $(\mathcal{T}_t)_{t\ge0}$
obtained under Assumption \ref{ass2.1}. Here we give their result.
\begin{thm} \label{thm2.3}[Theorem 4.4 in \cite{CF2}]
Suppose that there exists a positive self-adjoint operator $C$ in
$\ch$ with the following properties:
\begin{enumerate}
\item[(a)] The domain of the positive square root $C^{1/2}$ contains the domain $D(G)$ of $G$
  and $D(G)$ is a core for $C^{1/2}$ ,
 \item[(b)] the linear manifolds $L_l (D(G^2))$, $l \ge 1,$ are contained in the
domain of $C^{1/2}$,
\item [(c)] there exists a positive self-adjoint
operator $\Phi$, with $D(G)\subset D(\Phi^{1/2})$ such that, for
all $u\in D(G)$, we have
$$-2 \text{Re} \langle u,Gu \rangle  =\sum_{l=1}^\infty \|L_l u\|^2 =\|\Phi^{1/2}u\|^2,$$
 \item[(d)] $D(C)\subset D(\Phi)$, and for all $u\in D(C)$ we have
$\|\Phi^{1/2}u\|\le\|C^{1/2}u\|,$
\item[(e)]  there exists a positive constant $k$ such that
\begin{equation} \label{1.8-2}
2\text{Re} \langle C^{1/2} u, C^{1/2} G u \rangle +
\sum_{l=1}^\infty \|C^{1/2}L_l u\|^2\le k \| C^{1/2} u \|^2,
\end{equation}
for all $u\in D(G^2)$.
\end{enumerate}
Then the minimal q.d.s. $(\mathcal{T}_t)_{t\ge0}$ is conservative.
\end{thm}

\section{Conservative minimal quantum dynamical semigroups: Main results}
\quad Let  $\ch=L^2(\R^d)$ and $\cD=C_0^{\infty}(\R^d)$,
 the space of $C^\infty$-functions with compact support. We denote by $\partial_l=\frac{\partial}{\partial
  x_l}(\,l=1,2,...,d)$ differential operators with respect to the $l$-th coordinate
and $\partial_{lk}=\frac{\partial^2}{\partial x_k\partial
x_l}(\,l,k=1,2,...,d)$. For any measurable function $T$, we denote
the (distributional) derivative  $\frac{\partial T}{\partial x_l}$
by $(T)_l$, $\,l=1,2,...,d$. The Laplacian and  the gradient
operators are denoted by $\Delta$ and $\nabla,$ respectively.

Let a function (vector field) $W : \R^d \rightarrow \R^d$,
$W=(W_1, W_2,...,W_d),$ be given, where each component function
$W_l(x),\,\, l=1,2,...,d$, is a real valued twice differentiable
function on $\R^d$. We will denote
$$W^2=\sum_{l=1}^dW_l^2,\quad x^2=\sum_{l=1}^dx_l^2,\quad |x|=\Big(\sum_{l=1}^dx_l^2\Big)^{1/2}.$$
In the rest of this paper we suppose that $W$ satisfies the
following assumption.
\begin{ass}\label{ass3-0} The function  $W=(W_1,W_2,...,W_d)$
satisfies the following properties:
\begin{enumerate}
\item[{\bf (C-1)}] $W_l\in C^2(\R^d),\,l=1,2,...,d$,
\item[{\bf (C-2)}]
for any $\varepsilon\in(0,1)$ there exists a positive constant
$c(\varepsilon)$, depending on $\varepsilon$, such that
\begin{equation}
|(W_l)_k|\le\varepsilon W+c(\varepsilon) \label{3-0}
\end{equation}
for any $l,k=1,2,...,d$,
\item[{\bf (C-3)}]
there exist positive constants $c_1,\,c_2$ such that
\begin{equation}\label{3-0-1}
|(W_l)_{jk}|\le c_1|W|+ c_2,\,\,\,l,j,k=1,2,...,d.
\end{equation}
\end{enumerate}
\end{ass}
\begin{rem}\label{rem3}
(a)  By {\bf{(C-1)}}, $W_l^2\in L^2_{loc}(\R^d),\,l=1,2,...,d$.
Due to Theorem X.28 of \cite{RS},
 $-\Delta+W^2$ is essentially self adjoint on $\cD$.

(b)  The condition {\bf{(C-2)}} implies that for any
$\varepsilon\in(0,1)$ there exist positive constants
$c_1(\varepsilon)$ and $c_2(\varepsilon)$, depending on
$\varepsilon$, such that for any $l,k=1,2,...,d$ and  $u\in \cD$
\begin{eqnarray}
\|(W_l^2)_ku\|^2&\le&\varepsilon^2\|W^2u\|^2+c_1(\varepsilon)\|u\|^2,\label{3-16}\\
\|(W_l)_ku\|^2&\le&\varepsilon^2\|W^2u\|^2+c_2(\varepsilon)\|u\|^2.\label{3-16-1}
\end{eqnarray}

(c)  Using the fact that $|W|\le \frac12(\alpha W^2+\alpha^{-1}),
\alpha>0$, we get from (\ref{3-0-1}) that for any
$\varepsilon\in(0,1)$ there exist a positive constant
$c_3(\varepsilon^{-1})$, depending on $\varepsilon$
\begin{equation}\label{3-0*}
|(W_l)_{jk}|\le \varepsilon W^2+
c_3(\varepsilon^{-1}),\,\,\,l,j,k=1,2,...,d.
\end{equation}
\end{rem}
\begin{exam}\label{exam3-1}
Let $V :\R^d \rightarrow \R$ be the function (potential) given by
$$V(x)= \sum_{l=1}^{d}a_lx_l^{2n}+Q(x),$$ where
$a_l>0,\,\,\,l=1,2,...,d,$ and $Q(x)$ is a polynomial with degree
less than or equal to $2n-1$. Choose
$W=(W_1,W_2,...,W_d),\,W_l=\frac{1}{4}(V)_l,\,\,l=1,2,...,d.$ That
is, $W=\frac14 \nabla V.$ Then it is easy to check that for any
$l,k=1,2,...,d$,
\begin{eqnarray}\label{3-1}
|(W_l^2)_k(x)|&\le&\alpha_1|x|^{4n-3} +\beta_1,\\
W^2(x)&\ge&\alpha_2|x|^{4n-2}-\beta_2,\nonumber
\end{eqnarray}
for some positive constants $\alpha_1, \alpha_2, \beta_1$ and
$\beta_2.$ Notice that for any $\varepsilon>0$
\begin{eqnarray}\label{3-2}
|x|^{4n-3}&\le&\varepsilon|x|^{4n-2}\quad\text{if}\quad
|x|\ge\varepsilon^{-1},\\
|x|^{4n-3}&\le&\varepsilon^{-(4n-3)}\quad\text{if}\quad
|x|\le\varepsilon^{-1}.\nonumber
\end{eqnarray}
Combining (\ref{3-1}) and (\ref{3-2}), we get that the inequality
(\ref{3-0}) holds. The inequality (\ref{3-0-1}) can be checked
similarly. Thus $W$ satisfies Assumption \ref{ass3-0}.
\end{exam}

 Consider the operators $L_l,\, H,\, G_0$ and $G$ on a domain $\cD$
 \begin{eqnarray}
 &&L_lu=-(W_l+\partial_l)u,\,l=1,...,d,\quad L_l=0,\,l>d,\label{3-3.0}\\
 && Hu=\frac i2 \sum_{l=1}^d(W_l\partial_l+\partial_l W_l)u= \frac i2 \sum_{l=1}^d(2W_l\partial_l+(W_l)_l)u,\label{3-3}\\
&&G_0u=-\frac12 \sum_{l=1}^d L_l^*L_lu=-\frac12
\sum_{l=1}^d(W_l-\partial_l)(W_l+\partial_l)u\label{3-3.1}\\
&&\quad=-\frac12\big(-\Delta+W^2-\sum_{l=1}^d(W_l)_l\big)u,\nonumber\\
&&Gu=-iHu+G_0u.\label{3-3.2}
\end{eqnarray}
Clearly $H$ is a densely defined symmetric operator on $\cD$.
Recall that $-\Delta+W^2$ is essentially self adjoint on $\cD$.

\begin{lemm}\label{lemm3-0}
Suppose that $W=(W_1,W_2,...,W_d)$ satisfies Assumption
\ref{ass3-0}. Then the derivative $\sum_{l=1}^d(W_l)_l$ is
relatively $-\Delta+W^2$-bounded with relative bound less than $1$
on $\cD$. Moreover $-G_0$ is positive, essentially self adjoint on
$\cD$.
\end{lemm}
The proof of Lemma \ref{lemm3-0} will be given in Section 4. The
operator $G_0$ generates a strongly continuous contraction
semigroup on $\ch$. Since the adjoint operator $G^*$ of $G$ is
given by $G^*=iH+G_0$ on $\cD$,
 $G$ is closable. Denote by $G$ again its closure

 We consider the elliptic operator $\mathcal{L}$  on $B(\ch)$
 formally given by
\begin{eqnarray}\label{3-5}
\mathcal{L}(X)&=&i[H,X]-\frac12 \sum_{l=1}^d
L_l^*L_lX+\sum_{l=1}^d L_l^*XL_l-\frac12 \sum_{l=1}^d
XL_l^*L_l,\nonumber\\
 &=&G^*X+XG+\sum_{l=1}^d L_l^*XL_l,\quad X\in D(\mathcal{L}).
\end{eqnarray}
\begin{rem}\label{rem3-0}
In case that $[W_l,i\partial_l]$ is bounded on $\cD$ and $d=1$,
the elliptic operator $\mathcal{L}$ in (\ref{3-5}) was studied in
\cite{BK}. In this paper, we will remove the boundedness( see
(\ref{3-0})).
\end{rem}

 As mentioned in Introduction, we will construct the minimal
q.d.s. with the formal generator (\ref{3-5}) under Assumption
\ref{ass3-0}, and adding appropriate conditions (Assumption
\ref{ass3-1}), show the conservativity of the semigroup.

We state our main results. First let us introduce a proposition to
show that $G$ is the generator of a strongly continuous
contraction semigroup on $\ch$.
\begin{prop}\label{prop3.1-1}
Let $(A,D(A))$ be the generator of a strongly continuous
contraction semigroup on a Hilbert space $\ch$ and let $(B,D(B))$
be a symmetric operator on $\ch$. Assume that the following
properties hold:
\begin{enumerate}
\item[(a)] there is a dense set $D$ such that $D\subset D(A)\cap D(B)$
 and $D$ is a core for $A$,
\item[(b)] there are positive constants $a,\,b$ such that  the
bound
\begin{equation}\label{3-6}
\|Bu\|^2\le a^2\|Au\|^2+b^2\|u\|^2
\end{equation}
holds for any $u\in D$,
\item[(c)]
for any $\varepsilon>0$ there is a constant
$\tilde{c}(\varepsilon)>0$, depending on $\varepsilon$, such that
the bound
\begin{equation}\label{3-7}
\pm i\big(\langle Au,Bu\rangle-\langle
Bu,Au\rangle\big)\le\varepsilon\|Au\|^2+\tilde{c}(\varepsilon)\|u\|^2
\end{equation}
holds for any $u\in D.$
\end{enumerate}
Then for any $\alpha\in\R$ the operator $(A+i\alpha B, D(A))$
generates a strongly continuous contraction semigroup on $\ch$.
Moreover $D$ is a core for $A+i\alpha B.$
 \end{prop}

 Now consider the sesquilinear form $\mathcal{L}(X)$ on $\ch$ with
domain $\mathcal{D}\times \mathcal{D}$ given by
\begin{equation}\label{3-25}
\langle v,\mathcal{L}(X)u\rangle=\langle v,XGu \rangle+\langle
Gv,Xu \rangle+\sum_{l=1}^d\langle L_lv,XL_lu\rangle
\end{equation}
and the semigroup $\cT=(\cT_t)_{t\ge0}$ satisfying the equation
 \begin{equation}\label{3-27}
\langle v,\cT_t(X)u \rangle=\langle v,Xu \rangle+\int_0^t\langle
v,\mathcal{L}(\cT_s(X))u\rangle ds
\end{equation}
 for all $u,v\in \mathcal{D}$ and for all $X\in B(\ch)$.

\begin{thm}\label{thm3.2-1}
Suppose that $W=(W_1,W_2,...,W_d)$ satisfies Assumption
\ref{ass3-0}.
\begin{enumerate}
\item[(a)] The operator $G$ defined as in (\ref{3-3.2}) generates a
strongly continuous contraction semigroup on $\ch$. Moreover
$\cD=C_0^\infty(\R^d)$ is a core for $G.$
\item[(b)] There exists the minimal q.d.s. $\cT=(\cT_t)_{t\ge0}$ satisfying (\ref{3-27}).
\end{enumerate}
  \end{thm}

  Next, to show that the minimal q.d.s. $\cT=(\cT_t)_{t\ge0}$ is
conservative, let us introduce another assumption for
$W=(W_1,W_2,...,W_d)$.
\begin{ass}\label{ass3-1}
 There exists a constants $c_4\in\R$ such that
\begin{enumerate}
\item[\bf{(C-4)}] {$\big((W_l)_k\big)\ge-c_4$ in the
sense that for any complex numbers $\xi_1,\xi_2,...,\xi_d$,
$$\sum_{l,k=1}^d \overline{\xi_k}(W_l)_k\xi_l\ge -c_4\sum_{k=1}^d
|\xi_k|^2.$$}
\end{enumerate}
\end{ass}
\begin{rem}\label{rem3-0-1}
 Let $W=(W_1,W_2,...,W_d)$ be given as in Example
\ref{exam3-1}. Then {\bf (C-4)} means that $\text{Hess}\, V\ge
-c_5$, where $\text{Hess}\,\, V$ is the Hessian of $V$.
\end{rem}
\begin{thm}\label{thm3.3}
Suppose that $W=(W_1,W_2,...,W_d)$ satisfies Assumption
\ref{ass3-0} and Assumption \ref{ass3-1}. Then the minimal q.d.s.
$\cT=(\cT_t)_{t\ge0}$ satisfying (\ref{3-27}) is conservative.
 \end{thm}

 \section{Proofs of main results.}
 \quad In this section, we produce the proofs of Lemma \ref{lemm3-0}, Proposition \ref{prop3.1-1},
 Theorem \ref{thm3.2-1} and Theorem \ref{thm3.3}.
  We first give  the proof of Lemma \ref{lemm3-0}.

\vspace{0.5cm}
 {\bf Proof of Lemma \ref{lemm3-0}}: We compute that for $u\in\cD$
 \begin{eqnarray}\label{4.1}
\|(-\Delta+W^2)u\|^2&=&\|\Delta u\|^2+\|W^2u\|^2+2Re\langle-\Delta
u,W^2u\rangle\nonumber\\
     &=&\|\Delta u\|^2+\|W^2u\|^2\nonumber\\
     &&+2\sum_{l=1}^d \big(\langle\partial_l u,W^2\partial_l u\rangle
                 +Re\langle \partial_l u,(W^2)_l u\rangle\big)\nonumber\\
     &\ge&\|\Delta u\|^2+\|W^2u\|^2-2\sum_{l=1}^d \|\partial_l u\|\|(W^2)_lu\|\nonumber\\
    &\ge&\|\Delta u\|^2+\|W^2u\|^2-\sum_{l=1}^d\big(\|\partial_l u\|^2+\|(W^2)_lu\|^2\big).
 \end{eqnarray}
  Notice that for any $\tilde{\varepsilon}\in(0,1)$
 \begin{eqnarray}\label{4.2}
\sum_{l=1}^d\|\partial_l u\|^2&=&\langle -\Delta u,u\rangle\le\|\Delta u\|\|u\|\nonumber\\
&\le&\frac12\big(\tilde{\varepsilon}^2\|\Delta
u\|^2+\tilde{\varepsilon}^{-2}\|u\|^2\big).
 \end{eqnarray}
Choosing $\tilde{\varepsilon}$ sufficiently small, we conclude
from
 (\ref{4.1}), (\ref{4.2}) and the bound in (\ref{3-16}) that there
exist constants $b_1>1$ and $b_2>0$ such that
 \begin{equation}\label{4.4}
\|\Delta u\|^2+\|W^2u\|^2\le b_1\|(-\Delta+W^2)u\|^2+b_2\|u\|^2
 \end{equation}
 for any $u\in \cD$.

 Combining (\ref{4.4}) and (\ref{3-16-1}), and choosing $\varepsilon$ sufficiently small,  we obtain
 that
 \begin{equation}\label{4.6}
\|\sum_{l=1}^d(W_l)_lu\|^2\le b_3\|(-\Delta+W^2)u\|^2+b_4\|u\|^2
 \end{equation}
 for $u\in \cD$ and some $0<b_3<1$, $0<b_4$.
 This yields the proof of lemma.
$\square$

\vspace{0.5cm}
 {\bf Proof of Proposition \ref{prop3.1-1}}:  Replacing
$B$ by $a^{-1}B$, we may assume that $a=1$. It follows from
(\ref{3-7}) that for any $\gamma_1>0,\,\gamma_2>0$ and $u\in D$
\begin{eqnarray*}
&&\|(A+i\gamma_1B)u\|^2-\gamma_2^2\|Bu\|^2\\
&&\quad=\|Au\|^2+i\gamma_1\big(\langle Au,Bu\rangle-\langle
Bu,Au\rangle\big)+(\gamma_1^2-\gamma_2^2)\|Bu\|^2\\
&&\quad\ge(1-\gamma_1\varepsilon)\|Au\|^2+(\gamma_1^2-\gamma_2^2)\|Bu\|^2-
\gamma_1\tilde{c}(\varepsilon)\|u\|^2.
\end{eqnarray*}
By choosing $\varepsilon<\gamma_1^{-1}$, we conclude that for any
$0<\gamma_2\le\gamma_1$ and $u\in D$ the bound
\begin{equation}\label{3-8}
\gamma_2^2\|Bu\|^2\le\|(A+i\gamma_1B)u\|^2+\gamma_1\tilde{c}(\varepsilon)\|u\|^2
\end{equation}
holds.

Since $D$ is a core for $A$, the bound (\ref{3-6}) (with $a=1$)
holds for all $u\in D(A)$. Thus for any $0<\beta<1$, $\beta B$ is
relatively $A$-bounded with relative bound less than $1$. Since
$(B,D(B))$ is symmetric, it is dissipative. Therefore the operator
$(A+i\beta B,D(A))$ generates a strongly continuous contraction
semigroup on $\ch$(see Corollary 3.3 of [\cite {Paz}, Chap. 3].)
Moreover $D$ is a core for $A+i\beta B$ by (\ref{3-6})

The bound (\ref{3-8}) with $\gamma_1=\gamma_2=\beta$ implies that
for $0<\gamma<1$, $\beta\gamma B$ is relatively $A+i\beta
B$-bounded with relative bound less than $1$ and so
$(A+i\beta(1+\gamma)B,D(A))$ generates a strongly continuous
contraction semigroup and $D$ is a core for the operator. Since
$\beta\gamma<\gamma_2=\gamma_1=\beta(1+\gamma)$, the bound
(\ref{3-8}) implies that $(A+i\beta(1+2\gamma)B,D(A))$ generates a
strongly continuous contraction semigroup.

By using an induction argument, we conclude that for any
$\beta,\,\gamma\in(0,1)$ and $n=1,2,3,...$, the operator
$(A+i\beta(1+n\gamma)B,D(A))$ generates a strongly continuous
contraction semigroup and $D$ is a core for generator. For given
$\alpha>0$, one can choose $\beta,\gamma\in(0,1)$ and $n$ such
that $\alpha=\beta(1+n\gamma)$, and for given $\alpha<0$, $B$
replaces by $-B$. This completes the proof of the
theorem.$\square$

\vspace{0.5cm} In order to show that the operator $G$ defined as
in (\ref{3-3.2}) is a generator of a strongly continuous
contraction semigroup on $\ch$, we only need to check the
conditions of Proposition \ref{prop3.1-1}.

\vspace{0.5cm}
 {\bf Proof of Theorem \ref{thm3.2-1}}: (a) \quad To prove the
part (a) of theorem we apply Proposition \ref{prop3.1-1} for
$A=G_0,\,B=H$ and $D=\cD$. Clearly
 $H$ is a symmetric operator on $\cD$. By Lemma \ref{lemm3-0}, $G_0$ is negative, essential self-adjoint
on $\cD$, and so it generates a strongly continuous contraction
semigroup. Thus the condition (a) of Proposition \ref{prop3.1-1}
holds. Let us show the condition (b) of Proposition
\ref{prop3.1-1}. A direct computation yields that for $u\in \cD$
\begin{eqnarray*}\label{3-9}
\|Hu\|^2&=&\frac14\|\sum_{l=1}^d\big(2W_l\partial_l+(W_l)_l\big)u\|^2\\
        &\le&\frac{d}{4}\sum_{l=1}^d\|\big(2W_l\partial_l+(W_l)_l\big)u\|^2\\
         &\le&\frac{d}{2}\sum_{l=1}^d\big(4\|W_l\partial_l u\|^2+\|(W_l)_lu\|^2\big),
 \end{eqnarray*}
 and
 \begin{eqnarray*}\label{3-10}
\|W_l\partial_l u\|^2&=&\langle W_l^2\partial_l u,\partial_l u\rangle\nonumber\\
                 &=&\langle \partial_l W_l^2 u,\partial_l u\rangle-\langle (W_l^2)_l
                 u,\partial_l u\rangle\nonumber\\
                 &\le&\|W_l^2u\|\|\partial_l^2 u\|+\|(W_l^2)_l u\|\|\partial_l u\|\nonumber\\
                 &\le&\frac12 \big(\|W_l^2 u\|^2+\|\partial_l^2 u\|^2+\|(W_l^2)_l u\|^2+\|\partial_l u\|^2\big),
 \end{eqnarray*}
 which implies
 \begin{eqnarray}\label{3-11}
\|Hu\|^2&\le&d\sum_{l=1}^d \big(\|W_l^2 u\|^2+\|(W_l^2)_l
u\|^2+\frac12 \|(W_l)_lu\|^2\big)\nonumber\\
&&+d\sum_{l=1}^d \big(\|\partial_l^2 u\|^2+\|\partial_lu\|^2\big).
 \end{eqnarray}
Note that for $u\in \cD$
 \begin{eqnarray}
\sum_{l=1}^d \|W_l^2u\|^2&\le&\|W^2u\|^2,\label{4.3}\\
\sum_{l=1}^d\|\partial_l^2
u\|^2&\le&\|\sum_{l=1}^d\partial_l^2u\|^2=\|\Delta u\|^2,\nonumber
 \end{eqnarray}
  where we have used that for $l,k=1,2,...,d$
 \begin{equation*}
 \langle \partial_l^2 u,\partial_k^2 u\rangle=\langle\partial_{lk}
 u,\partial_{kl}u\rangle\ge0.
 \end{equation*}
  Applying (\ref{4.2}), (\ref{4.3}) and  (\ref{3-16}) into (\ref{3-11}), we get
 that there exist constants $a_1>d$ and $a_2>0$ such that for any $u\in \cD$
\begin{equation}\label{3-18}
\|Hu\|^2\le a_1(\|\Delta u\|^2+\|W^2u\|^2)+a_2\|u\|^2.
 \end{equation}

 On the other hand, for any $\varepsilon\in(0,1)$ and $u\in\cD$, we have
\begin{eqnarray}\label{3-12}
\|G_0u\|^2&=&\frac14\|\big(-\Delta+W^2-\sum_{l=1}^d(W_l)_l\big)u\|^2\nonumber\\
          &\ge&\frac14\Big(\|(-\Delta+W^2)u\|-\|\sum_{l=1}^d(W_l)_lu\|\Big)^2\nonumber\\
          &\ge&\frac14\Big((1-\varepsilon)\|(-\Delta+W^2)u\|^2
          +(1-\varepsilon^{-1})\|\sum_{l=1}^d(W_l)_lu\|^2\Big)\nonumber\\
          &\ge&\frac14\Big((1-\varepsilon)\|(-\Delta+W^2)u\|^2
          -\varepsilon^{-1}\sum_{l=1}^d\|(W_l)_lu\|^2\Big).
 \end{eqnarray}
Substituting (\ref{4.1}) into (\ref{3-12}), we have
 \begin{eqnarray}\label{3-14}
\|G_0 u\|^2&\ge&\frac14\big(1-\varepsilon)\big(\|\Delta
u\|^2+\|W^2u\|^2\big)\nonumber\\
&&-\frac14\sum_{l=1}^d\big(\|\partial_l
u\|^2+\|(W^2)_lu\|^2+\varepsilon^{-1}\|(W_l)_lu\|^2\big).
 \end{eqnarray}
Choosing $\varepsilon$ sufficiently small, we conclude from
(\ref{3-14}), (\ref{4.2}) and the bound in (\ref{3-16}) that there
exist constants $a_3>4$ and $a_4>0$ such that
 \begin{equation}\label{3-17}
\|\Delta u\|^2+\|W^2u\|^2\le a_3\|G_0u\|^2+a_4\|u\|^2
 \end{equation}
 for any $u\in \cD$.
  Combining (\ref{3-17}) and (\ref{3-18}), we obtain
 that
 \begin{equation}\label{3-19}
\|Hu\|^2\le a_5\|G_0 u\|^2+a_6\|u\|^2,\,\,u\in\cD
 \end{equation}
 for some $a_5>4d$ and $a_6>0$. This proves the inequality
 (\ref{3-6}).

Next we consider the commutator estimate in (\ref{3-7}). Recall
that
 \begin{eqnarray}\label{3-20}
&&A=G_0=-\frac12 \sum_{l=1}^d L_l^*L_l=-\frac12
(-\Delta+W^2-\sum_{l=1}^d(W_l)_l),\nonumber
 \\&&B=H=\frac i2\sum_{l=1}^d(W_l\partial_l+\partial_l W_l).
\end{eqnarray}
We can write that
\begin{eqnarray}\label{3-21}
\pm i[-\frac12\sum_{l=1}^dL_l^*L_l,H]
&=&\pm\frac14[-\Delta+W^2-\sum_{l=1}^d(W_l)_l,
                                       \sum_{k=1}^d(W_k\partial_k+\partial_k W_k)]\nonumber\\
         &=&\pm\frac14 \sum_{k=1}^d[-\Delta,W_k\partial_k+\partial_k W_k]\nonumber\\
         &&\pm\frac14\sum_{k=1}^d[W^2-\sum_{l=1}^d(W_l)_l,W_k\partial_k+\partial_k W_k].
         \end{eqnarray}
 Notice that
 \begin{eqnarray}\label{3-21*}
 &&[-\Delta,W_k\partial_k+\partial_k W_k]=[-\Delta,W_k]\partial_k+\partial_k[-\Delta,W_k]\\
 &&\quad=-\sum_{l=1}^d\big(\partial_l(W_k)_l\partial_k+(W_k)_l\partial_{kl}
 +\partial_{lk}(W_k)_l+\partial_k(W_k)_l\partial_l\big)\nonumber\\
 &&\quad=-\sum_{l=1}^d\big((W_k)_{ll}\partial_k+
 2(W_k)_l\partial_{kl}+2\partial_{lk}(W_k)_l-\partial_k(W_k)_{ll})\big)\nonumber
 \end{eqnarray}
 and
 \begin{eqnarray}\label{3-22*}
&&[W^2-\sum_{l=1}^d(W_l)_l,W_k\partial_k+\partial_k W_k]\\
&&\quad=[W^2,W_k\partial_k+\partial_kW_k]-\sum_{l=1}^d[(W_l)_l,W_k\partial_k+\partial_kW_k]\nonumber\\
 &&\quad=-2W_k\big((W^2)_k-\sum_{l=1}^d(W_l)_{lk}\big)\nonumber\\
 &&\quad=-2\sum_{l=1}^dW_k\big((W_l^2)_k-(W_l)_{lk}\big)\nonumber
         \end{eqnarray}
as bilinear forms on $\cD$. Substituting (\ref{3-21*}) and
(\ref{3-22*}) into (\ref{3-21}), we obtain that for $u\in\cD$
\begin{eqnarray}\label{3-21-1}
 \pm  \langle u,i[-\frac12 \sum_{l=1}^d L_l^*L_l,H]u\rangle
 &=&\mp\sum_{l,k=1}^d Re \big(\langle (W_k)_l u,\partial_{kl} u\rangle
 +\frac12\langle (W_k)_{ll} u, \partial_k u\rangle\big)\nonumber\\
 &&\mp\frac12 \sum_{l,k=1}^d\langle W_ku, \big((W_l^2)_k-(W_l)_{lk}\big)u \rangle.
 \end{eqnarray}
Notice that for $\varepsilon\in(0,1)$ and $u\in \cD$ ,
\begin{eqnarray}\label{3-22}
\sum_{l,k=1}^d|\langle (W_k)_lu,\partial_{kl}u\rangle|
&\le&\sum_{l,k=1}^d\|(W_k)_lu\|\|\partial_{kl}u\|\nonumber\\
&\le&\sum_{l,k=1}^d\frac12\big(\varepsilon\|\partial_{kl}u\|^2+\varepsilon^{-1}\|(W_k)_lu\|^2\big)\nonumber\\
&\le&\sum_{l,k=1}^d\frac12\big(\varepsilon\langle\partial_l^2u,\partial_k^2
u\rangle+\varepsilon\|W^2u\|^2+b_1\|u\|^2\big)\nonumber\\
&=&\frac12\big(\varepsilon\|\Delta
u\|^2+d^2\varepsilon\|W^2u\|^2+b_1d^2\|u\|^2\big)
\end{eqnarray}
for some constant $b_1>0.$
 Here we have used (\ref{3-16-1}) in third inequality.
 Similarly,  for $\varepsilon\in(0,1)$ and $u\in \cD$ we get from
 (\ref{3-0*}) and (\ref{4.2}) that
\begin{eqnarray}\label{3-22-1}
\sum_{l,k=1}^d|\langle (W_k)_{ll}u,\partial_ku\rangle|
&\le&\sum_{l,k=1}^d\frac12\big(\|(W_k)_{ll}u\|^2+\|\partial_ku\|^2\big)\nonumber\\
&\le&\frac12\big(\varepsilon d\|\Delta
u\|^2+d^2\varepsilon\|W^2u\|^2+b_2\|u\|^2\big),
\end{eqnarray}
 and also by (\ref{3-16}) and (\ref{3-0*}),
\begin{eqnarray}\label{3-22-0}
&& \sum_{l,k=1}^d|\langle W_ku,\big((W_l^2)_k-(W_l)_{lk}\big)u\rangle|\\
&&\quad\le\sum_{l,k=1}^d\frac12\big(\varepsilon\|W_ku\|^2+\varepsilon^{-1}\big(\|(W_l^2)_ku\|^2
       +\|(W_l)_{lk}u\|^2\big)\big)\nonumber\\
 &&\quad\le\varepsilon b_3\|W^2u\|^2+b_4\|u\|^2\nonumber
\end{eqnarray}
for some constants $b_2, \,b_4$ depending on $\varepsilon$ and
$b_3>0$, where we have used
\begin{equation*}
\sum_{k=1}^d\|W_ku\|^2=\sum_{k=1}^d\langle u,W_k^2
u\rangle=\langle u,W^2 u\rangle\le\frac12(\|u\|^2+\|W^2u\|^2).
\end{equation*}
 Then substituting
(\ref{3-22}), (\ref{3-22-1}) and (\ref{3-22-0}) into
(\ref{3-21-1}),  one has that  $\tilde{\varepsilon}\in(0,1)$ and
some $b_5>0$
\begin{equation}\label{3-23}
\pm\langle u,i[-\frac12\sum_{l=1}^dL_l^*L_l,-H]u\rangle
\le\tilde{\varepsilon}(\|\Delta u\|^2+\|W^2u\|^2)+b_5\|u\|^2.
 \end{equation}
 Here we can choose  $\tilde{\varepsilon}$  as small as possible.
Hence two inequalities (\ref{3-17}) and (\ref{3-23}) produces that
for any $\varepsilon'>$ there is a constant $b_6$, depending on
$\varepsilon'$, such that the bound
\begin{equation}\label{3-23-3}
\pm\langle u,i[-\frac12\sum_{l=1}^dL_l^*L_l,-H]u\rangle
\le\varepsilon'\|G_0u\|^2+b_6\|u\|^2.
 \end{equation}
  The part (a) of the proof is completed.

(b)\quad By (a), $G$ generates a strongly continuous contraction
semigroup on $\ch$ and $\cD=C_0^\infty(\R^d)$ is a core for $G.$
We get from (\ref{3-3.1}) and (\ref{3-3.2}) that we have
\begin{equation}\label{3-26}
\langle v,Gu \rangle+\langle Gv,u \rangle+\sum_{l=1}^d\langle
L_lv,L_lu\rangle=0
\end{equation}
for all $u,v\in \mathcal{D}$. Thus $G$ and $L_l,\,l=1,2,...,d,$
satisfy the condition $(\ref{2.1})$ on $\mathcal{D}$, a core for
$G$, and so Assumption \ref{ass2.1} is satisfied. Therefore, as
mentioned in Section 2, by the iterations, we can construct a
minimal q.d.s. $\cT=(\cT_t)_{t\ge0}$ satisfying the equation
 (\ref{3-27}).
$\square$

\vspace{0.5cm} {\bf Proof of Theorem \ref{thm3.3}}:  Applying
Theorem \ref{thm2.3}, we show that the minimal q.d.s. is
conservative.
 Let us choose the operator $C$
 \begin{eqnarray}\label{3-30}
 &&C=-2G_0=\sum_{l=1}^dL_l^*L_l=-\Delta+W^2-\sum_{l=1}^d(W_l)_l,\\
 && D(C)=\{u\in L^2(\R^d)|\text{the distribution}\,\, Cu\in L^2(\R^d)\}\nonumber.
 \end{eqnarray}
 Recall that $\cD$ is a core for $C$. We have that as bilinear forms on $\cD$
\begin{eqnarray}
G^* G &=& (i H + G_0)(-i H + G_0) \nonumber \\ &=& H^2 + G_0 ^2 +
i [H, G_0] \nonumber \\ & \ge & G_0^2 + i [H, G_0 ]. \label{3-31}
\end{eqnarray}
  It follows from (\ref{3-31}) and (\ref{3-23-3}) that
we have
\begin{equation}\label{3-32}
\|G_0u\|^2 \le a\|Gu\|^2+b\|u\|^2,\quad u\in\cD
\end{equation}
for some $a,\,b>0$. Using the relations (\ref{3-30}), (\ref{3-32})
and the fact that $-iH$ is relatively bounded perturbation of
$G_0$, we obtain  that $G$ and $C$ are relatively bounded
 with respect to each other and so $D(G)$ = $D(C)$.

 We will check that the operator $C$ satisfies Theorem
 \ref{thm2.3}.  Hypothesis (a) and (b) of Theorem
   \ref{thm2.3} are trivially fulfilled.
 To check the condition (e) of Theorem
   \ref{thm2.3}, we estimate
\begin{equation}\label{3-33}
 C G+ G^* C + \sum_{l=1}^d L_l^* C L_l   = i[H,C] +\frac
12 \sum_{l=1}^d\big( L_l^* [C, L_l] + (L_l^*[C, L_l])^* \big)
\end{equation}
as bilinear forms on $\cD.$

 We obtain from (\ref{3-21-1}) and $C=\sum_{l=1}^d L_l^* L_l$ that
 \begin{eqnarray}\label{3-34}
i[H,C]&=&-\sum_{l,k=1}^d\big((W_k)_l\partial_{kl}+\partial_{lk}(W_k)_l+\frac12\big((W_k)_{ll}\partial_k-\partial_k(W_k)_{ll}\big)\big)\nonumber\\
&&-\sum_{l,k=1}^d W_k\Big((W_l^2)_k-(W_l)_{lk}\Big)\nonumber\\
&=&-\sum_{l,k=1}^d\big(2\partial_l(W_k)_l\partial_k+\partial_l(W_k)_{lk}
-\frac12\big((W_k)_{ll}\partial_k+\partial_k(W_k)_{ll}\big)\big)\\
&&-\sum_{l,k=1}^d\big(2W_kW_l(W_l)_k-W_k(W_l)_{lk}\big)\nonumber
\end{eqnarray}
as bilinear forms on $\cD$.
 On the other hand, we have
\begin{eqnarray*}
[C,L_l]&=&[-\Delta+W^2-\sum_{k=1}^d(W_k)_k,W_l+\partial_l] \\
        &=&-[\Delta,W_l]+[W^2,\partial_l]-\sum_{k=1}^d[(W_k)_k,\partial_l] \\
         &=&\sum_{k=1}^d\big(-\partial_k(W_l)_k-(W_l)_k\partial_k-2W_k(W_k)_l+(W_k)_{kl}\big)\\
         &=&\sum_{k=1}^d\big(-2\partial_k(W_l)_k+(W_l)_{kk}-2W_k(W_k)_l+(W_k)_{kl}\big),
\end{eqnarray*}
 which implies
\begin{eqnarray*}
&&\frac12\sum_{l=1}^d(L_l^*[C,L_l]+(L_l^*[C,L_l])^*)\\
&&\quad=\frac12\sum_{l,k=1}^d\big\{(W_l-\partial_l)(-2\partial_k(W_l)_k+U(k,l))+(2(W_l)_k\partial_k+U(k,l))(W_l+\partial_l)\big\}\\
&&\quad=\sum_{l,k=1}^d\big\{\big((W_l)_k\partial_kW_l-W_l\partial_k(W_l)_k\big)+W_lU(k,l)\big\}\\
&&\quad\,\,\,+\sum_{l,k=1}^d\big\{\big(\partial_{kl}(W_l)_k+(W_l)_k\partial_{lk}\big)-\frac12[\partial_l,U(k,l)]\big\}
\end{eqnarray*}
as bilinear forms on $\cD$, where
$U(k,l)=(W_l)_{kk}-2W_k(W_k)_l+(W_k)_{kl}$. Notice that
$$[\partial_l,U(k,l)]=[\partial_l,(W_l)_{kk}+(W_k)_{kl}]-2((W_k)_l)^2-2W_k(W_k)_{ll}$$
as bilinear forms on $\cD$. Thus we have
\begin{eqnarray}\label{3-35}
&&\frac12\sum_{l=1}^d(L_l^*[C,L_l]+(L_l^*[C,L_l])^*)\nonumber\\
&&=\sum_{l,k=1}^d\big\{\big(((W_l)_k)^2-W_l(W_l)_{kk}\big)+W_l\big((W_l)_{kk}-2W_k(W_k)_l+(W_k)_{kl}\big)\big\}\nonumber\\
&&\quad+\sum_{l,k=1}^d\big(2\partial_k(W_l)_k\partial_l+\partial_k(W_l)_{kl}-(W_l)_{kk}\partial_l\big)\nonumber\\
&&\quad+\sum_{l,k=1}^d\big(-\frac12[\partial_l,(W_l)_{kk}+(W_k)_{kl}]+((W_k)_l)^2+W_k(W_k)_{ll}
\big)\nonumber\\
&&=\sum_{l,k=1}^d\big\{2\big(((W_l)_k)^2+\partial_k(W_l)_k\partial_l-W_lW_k(W_k)_l\big)
+\big(W_l(W_k)_{kl}+W_k(W_k)_{ll}\big)\big\}\nonumber\\
&&\quad+\sum_{l,k=1}^d\big(\big(\partial_k(W_l)_{kl}-(W_l)_{kk}\partial_l\big)
-\frac12[\partial_l,(W_l)_{kk}+(W_k)_{kl}]\big),
\end{eqnarray}
as bilinear forms on $\cD$.
 Exchanging $l$ and $k$ in (\ref{3-34}), and substituting (\ref{3-34}) and (\ref{3-35}) into
(\ref{3-33}),  one has
\begin{eqnarray}\label{3-36-1}
 &&C G+ G^* C + \sum_{l=1}^d L_l^* C L_l\\
 &&\quad =\sum_{l,k=1}^d\big(-4W_lW_k(W_k)_l-\frac12[\partial_l,(W_k)_{kl}]\big)\nonumber\\
&&\quad\,\,+\sum_{l,k=1}^d\big(2((W_k)_l)^2+2W_l(W_k)_{kl}+W_k(W_k)_{ll}\big)\nonumber
\end{eqnarray}
as bilinear forms on $\cD.$
 By ({\bf C-4}), we get that for $u\in\cD$
\begin{eqnarray}\label{3-37}
-4\langle u,\sum_{l,k=1}^dW_l(W_k)_lW_k u\rangle
&=&-4\sum_{l=1}^d\langle W_l u,\sum_{k=1}^d(W_k)_lW_k u\rangle\nonumber\\
&\le&4c_5\sum_{l=1}^d\langle W_l u,W_l u\rangle=4c_5\langle
u,W^2u\rangle.
\end{eqnarray}
And it follows from ({\bf C-2}) and ({\bf C-3})  that
\begin{eqnarray}\label{3-38}
|2((W_k)_l)^2+2W_l(W_k)_{kl}+W_k(W_k)_{ll}|&\le& b_1W^2+b_2,\\
\sum_{l,k=1}^d|\langle u,\partial_l(W_k)_{kl}u\rangle|
&\le&\sum_{l,k=1}^d\frac12\big(\|\partial_lu\|^2+\|(W_k)_{kl}u\|^2\big)\nonumber\\
&\le&\frac12\big(d\langle u,-\Delta u\rangle+d^2\langle
u,(b_3W^2+b_4)u\rangle\big),\nonumber
\end{eqnarray}
for $u\in\cD$ and some positive constants $b_i,\,i=1,2,3,4$.
Applying (\ref{3-37}) and (\ref{3-38}) into (\ref{3-36-1}), we
have
  \begin{equation*}
2 Re \langle Cu , G u \rangle + \sum_{l=1}^d\langle L_lu , CL_l u
\rangle\le b_5 \langle u,(-\Delta+W^2)u\rangle+b_6\langle u
,u\rangle,\quad u\in \cD
\end{equation*}
for some $b_5, b_6>0$.  By ({\bf C-2}), for $u\in \cD$ and
$\varepsilon\in(0,1)$ we have
\begin{eqnarray*}
\langle u,Cu\rangle&=&\langle
u(-\Delta+W^2)u\rangle-\sum_{l=1}^d\langle u
,(W_l)_lu\rangle\\
&\ge& \langle u,(-\Delta+W^2)u\rangle-\varepsilon\langle u
,W^2u\rangle-b_7\|u\|^2 \\
&\ge& (1-\varepsilon)\langle u,(-\Delta+W^2)u\rangle-b_7\|u\|^2,
\end{eqnarray*}
for some constant $b_7>0.$ Thus for $u\in \cD$ and some $b_8,
b_9>0$
\begin{equation*}
2 Re \langle Cu , G u \rangle + \sum_{l=1}^d\langle L_lu , CL_l u
\rangle\le b_8 \langle u,Cu\rangle+b_9\langle u ,u\rangle.
\end{equation*}
Redefine $C=\sum_{l=1}^d L_l^* L_l+\frac{b_9}{b_8}$, then by
(\ref{3-26}) we have
      \begin{equation}\label{3-41}
      2 Re \langle Cu , G u \rangle + \sum_{l=1}^d\langle L_lu , CL_l u
\rangle \le b_8 \langle u,Cu\rangle,\,u\in\cD.
       \end{equation}
        We want to extend the inequality (\ref{3-41})  to
the domain $D(G)$. Since $G$ and $C$ are relatively bounded
 with respect to each other, there exists a sequence $\{u_n
\}$ of elements of $\cD$ such that $$ \lim_{n \to \infty} u_n =u,
\,\,\lim_{n \to \infty} Cu_n =Cu, \,\,\lim_{n \to \infty} Gu_n
=Gu,\quad u\in D(G). $$
 Then the relation (\ref{3-41}) implies that $ \{C^{1/2} L_l u_n \}_{n\ge 1}$ is a Cauchy sequence. Therefore it is
convergent and it is easy to deduce that (\ref{3-41}) holds for $u
\in D(G)$.

Note that $\Phi= \sum_{l=1}^d L_l^* L_l\le
C(=\Phi+\frac{b_9}{b_8})$ as bilinear forms on $\cD$. Hence the
conditions (c), (d) of Theorem
   \ref{thm2.3} also hold and the minimal
q.d.s. is conservative.$\square$

\vspace{0.2cm} \noindent {\bf Acknowledgements :} This work was
supported by Korea Research Foundation  Grant (KRF-2003-005-00010,
KRF-2003-005-C00011).

\end{document}